\documentclass{article}
\usepackage{authblk}
\usepackage{graphicx} 
\usepackage[left=1.5in, right=1.5in, top=1.5in, bottom=1.5in]{geometry}
\usepackage[natbibapa]{apacite}
\usepackage{xcolor}
\usepackage{adjustbox}
\usepackage{float}
\usepackage{subcaption}
\usepackage{hyperref}

\title{Academic Freedom and International Research Collaboration: A Longitudinal Analysis of Global Network Evolution}

\author[1]{Travis A. Whetsell}

\affil[1]{School of Public Policy, Georgia Institute of Technology, Atlanta, Georgia, travis.whetsell@gatech.edu, ORCID: 0000-0001-5395-4754, corresponding author}

\author[2]{Jen Sidorova}

\affil[2]{School of Public Policy, Georgia Institute of Technology, Atlanta, Georgia, esidorova3@gatech.edu, ORCID: 0000-0002-0924-535X}

\date{\today}

\begin{document}

\maketitle

\begin{abstract}

The topic of academic freedom has come to the fore as nations around the world experience a wave of democratic backsliding. Institutions of higher education are often targets of autocrats who seek to suppress intellectual sources of social and political resistance. At the same time, international collaboration in scientific research continues unabated, and the network of global science grows larger and denser every year. This research analyzes the effects of academic freedom on international research collaboration (IRC) in a sample of 166 countries. Global international collaboration data are drawn from articles in Web of Science across a 30-year time frame (1993-2022) and are used to construct three separate IRC networks in science and technology (S\&T), social sciences (SocSci), and arts and humanities (A\&H). The Academic Freedom Index, covering the same time frame, is drawn from the Varieties of Democracy Project, as are numerous country-level control variables. Stochastic actor-oriented models (SAOM) are used to analyze the networks. The results show positive significant estimates for both direct effects and homophily effects of academic freedom on network evolution. These effects appear to increase in strength moving from the S\&T network, to the SocSci network, and appear strongest in the A\&H network. However, tests of temporal heterogeneity show a significant decline in the relevance of academic freedom in the most recent period. 

\end{abstract}

\section*{\small \centering Acknowledgements} {\small RSiena models presented here were initially developed at the 14th Winter School on Longitudinal Social Network Analysis at the University of Groningen. This research was presented at the INSNA Sunbelt 2024 conference. We would like to thank John Walsh, Diana Hicks, and Caroline Wagner for comments on earlier versions of this paper. An earlier version of this paper has an arxiv preprint arXiv:2407.03968.}

\newpage

\section{Introduction}

Scholars have long believed that factors of national governance, such as openness and liberalism, are requisites to the flourishing of scientific inquiry \citep{polanyi1947foundations,merton1973sociology, popper2012open}. However, assumptions underlying this belief have yet to be fully analyzed using the tools of modern computational social science. Recent developments in both the empirical measurement of academic freedom \citep{Spannagel2023} and scientometric methodology \citep{Fortunato2018} have made it possible to conduct global scale tests of theory. 

Exploring the relationships between factors of national governance and the dynamics of scientific research has become all the more relevant as the world undergoes a wave of democratic backsliding \citep{luhrmann2019}. Institutions of higher education are often targets of autocrats who seek to suppress intellectual sources of social and political resistance. Simultaneously, international collaboration in scientific research has reached an all-time high, and the network of global science grows larger and denser every year. Recent research has begun to connect factors of national governance with processes and outcomes of scientific research \citep{Whetsell2021,Whetsell2023}. 

This research analyzes the effects of academic freedom on international research collaboration (IRC) in a sample of 166 countries assembled within a global network of science. The Academic Freedom Index \citep{Spannagel2023} is drawn from the Varieties of Democracy Project \citep{Coppedge2011}. Global international collaboration data are drawn from Web of Science (WoS) across a 30-year time frame (1993-2022). From these IRC data, longitudinal networks are constructed on three broad subject domains, including science and technology, social sciences, and arts and humanities. Lastly, the time frame is segmented into three consecutive decades to determine whether the significance of academic freedom has declined in recent years.

Stochastic actor-oriented models (SAOM) are used to analyze networks, implemented in the \texttt{R} package \texttt{RSiena} \citep{Snijders2023a}. SAOMs are inferential network models useful for accounting for dependencies in longitudinal network data \cite{cranmer2020inferential}. Endogenous network control variables are included in the model, as are exogenous country level factors, including geographic distance, population size, urbanization, and GDP. The results show positive significant estimates for a direct linear effect and a homophily effect of academic freedom on tie creation and maintenance over time in the global IRC network. These effects vary in theoretically anticipated ways across three domains of research. However, the results also show significant declines in the effect of academic freedom on IRC network evolution in recent years. 

\section{Literature Review}

The literature used to generate hypotheses for this research project represents two broad subject areas. First, a relatively small set of recent articles orbiting political science analyzes academic freedom as an element of democratic governance. Second, the relevant literature on international research collaboration has largely developed in journals of informetrics, scientometrics, and research policy. While not completely unrelated, these literatures tend to ignore one another. At the same time, there has been a surprising lack of attention to country level political factors in international research policy. Thus, the following literature review combines insights from both subject areas to test theory. 

\subsection{Academic Freedom}

Historically, the concept of academic freedom arose from the fundamental tensions among the university as an autonomous institution of society, the influence of organized religion, and the authority of the modern nation-state \citep{Levine2021}. The classic allegory illustrating these tensions is Plato's Trial of Socrates, where the quintessential educator is put to death by Athens for impiety and corruption of the youth. While the idea of intellectual freedom has a long history rooted in ancient Greece, the Renaissance, and the Age of Reason, development and institutionalization of academic freedom is a more recent phenomenon associated with the modern university system \citep{fuchs1963academic}. In the 19th century German context \textit{Wissenschaft}, the systematic pursuit of truth encompassing all disciplines, was to be the mission of the university, and academic freedom (\textit{Akademische Freiheit}) provided the means to achieve this mission understood as the liberty to teach (\textit{Lehrfreiheit}) and to learn (\textit{Lernfreiheit})\citep{hart1874german}. Many prominent American scholars studied in Germany during the mid-19th century and returned to the United States with these new concepts influencing the development of the American university system \citep{rockwell1950academic}. 

In 1915 the American Association of University Professors (AAUP) was formed in response to numerous infringements of academic freedom. The AAUP expanded the concept to include not only teaching and research but also intramural and extramural expression. Academic freedom has now diffused across the international system, becoming a foundational principle in the university systems of many countries \citep{altbach2007academic}. However, the diffusion of academic freedom has not been completely uniform. Research shows a cyclical pattern in the growth and decline of academic freedom in the 1940s, the late-1970s, and in the 2010s \citep{lott2023academic}. More recently, several studies have documented numerous instances of infringement on academic freedom across the world \citep{maassen2023state,garry2023threats,darian2024knowledge}. 

Despite the ups and downs in academic freedom, research continues to show that it plays a vital role in ensuring quality in higher education, generating scientific and technological advancements, and preserving democratic values in countries around the world \citep{giroux2002neoliberalism, clark1983contradictions, Enders2007, karran2009academic}. Recent literature has reinforced and broadened previous findings, underscoring academic freedom's critical role in fostering innovation, facilitating knowledge exchange, and bolstering societal welfare. For instance, \cite{fernandez2024science} observed that academic freedom significantly improves the quality of STEM publications. \cite{vogtle2024does} find that countries with stronger academic freedom draw more international students and have a positive influence on global student mobility. \cite{posso2023social} underscored the contribution of academic freedom to social R\&D, societal conditions, and the reduction in inequality. Recent research shows that academic freedom is associated with greater technological innovation \citep{audretsch2024academic}. 

A separate strand of literature has investigated the relationship between academic freedom and institutions of political governance more broadly. \cite{pernia2023academic} and \cite{lerch2024social} find that academic freedom and democracy are highly correlated, and \cite{berggren2022political} shows that the presence of democratic institutions predicts academic freedom. \cite{spannagel2023introducing} analyzed academic freedom in national constitutions and highlights the heterogeneity of regional and ideological models that formed and spread over time. \cite{kinzelbachacademic}'s investigation into academic freedom in Asia from 1900 to 2021 revealed that shifts in academic freedom are explained by power transitions, including processes of democratization or autocratization. \cite{karran2007academic} highlighted the varying degrees of constitutional protection of academic freedom among European countries and identified a trend towards managerialism that originates in the UK and is detrimental to academic freedom. 

Conversely, \cite{pelke2023academic} examined how academic freedom slows the onset of autocratization, suggesting that increased academic freedom fosters greater support for democratic institutions. \cite{kumaraeffect} found that current levels of electoral democracy correlate with previous levels of academic freedom in a sample of Asian countries. \cite{kariuki2022attacks} noted that African nations with greater academic freedom during democratic transitions later achieved higher levels of democracy. \cite{kratou2022impact} found a positive association between the quality of post-transition elections and preceding levels of academic freedom. \cite{enyedi2016populist} discovered that the erosion of academic freedom in Hungary coincided with the rise of electoral autocracy. 

Although the literature generally supports a link between democratic institutions and academic freedom, its effect on international collaborations remains ambiguous. While \cite{enyedi2016populist} suggests that the autocratization of nations could impede scientists' self-organization across borders, \cite{chankseliani2021big} observe that Post-Soviet researchers actively engage in international networks, resulting in a significantly co-authored output. This dynamic, attributed to a lack of resources and institutional support, implies that while democratic institutions naturally encourage collaboration, autocratic systems may indirectly force scholars toward international collaboration in search of independent research opportunities. Recent studies highlight the persistence of a high level of international academic collaboration even in autocratic regimes like China \citep{zhang2017china, niu2014network}. This paradox suggests that while democratic environments foster academic freedom, autocratic pressures may also drive scholars towards extensive international cooperation, possibly as a strategy to overcome restrictive national conditions. This requires further empirical investigation in order to better understand the nuanced impacts of political regimes on academic practices. 

Finally, there are reasons to believe that academic freedom can vary by academic discipline. Certain disciplines may be more or less influenced by external pressures from governments, religion, industry, or even overbearing university administrative systems. Academic freedom can be restricted by industry influences, such as funding, career opportunities, collaboration and conference sponsorship \citep{ebell2021towards,kirkham2022industrial,behrens2001unintended}. \cite{steinmetz2018scientific} in analysis of power relations within American Universities classifies heterogeneous, practical and applied sciences such as applied math, computer science, public policy and medicine as less autonomous from a larger administrative cloud that includes university administration, trustees, and state legislators (in public universities), but theoretical sciences such as math or physics may have greater autonomy. Thus, in domains of research that are more politically, religiously, or culturally salient, e.g. social sciences or the arts and humanities, international collaboration may be more greatly affected by academic freedom.

\subsection{International Research Collaboration}

International research collaboration (IRC) has expanded rapidly from its early 20th century origins, propelled by the global networking activities of scientists \citep{price1963little, castelvecchi2015physics, wagner2009new, wagner2017growth, wagner2015continuing}. This expansion is driven by technological advances that reduce collaboration costs, as well as the increasing complexity and specialization occurring within scientific fields \citep{landry1998impact, huang2015study}. The rise of multidisciplinary projects and a general shift in institutional norms to promote collaborative research have further accelerated IRC's growth \citep{wray2002epistemic, wagner2011approaches}. The change in institutional norms accelerated the impact of IRC on productivity and its power to enable researchers to achieve broader educational and professional objectives through collaboration \citep{defazio2009funding}. Scientific productivity and collaboration have been found to mutually enhance each other, with collaboration correlating with higher scientific output and greater citation impact \citep{fox2021being, glanzel2002distributional, van1997science, robinson2019many, leydesdorff2019relative}. However, there is a bidirectional relationship, as more productive scientists tend to engage in more collaboration \citep{abramo2011relationship, sugimoto2017scientists}. While this literature enhances our understanding of driving factors in IRC proliferation, it tends to avoid the analysis of internal country factors in relation to IRC.

More recent research has revealed some key patterns and properties of collaborative networks. \cite{fitzgerald2021academia} show that traditional collaboration patterns based on historical and geopolitical lines have shifted towards more strategic, region-based partnerships. Despite these changes, the dominance of scholars from Anglophone countries remains largely unchanged, indicating persistent imbalances in global research leadership \citep{fu2022evolving}. Additionally, new trends such as "Shared Heritage IRC" are emerging \citep{gok2024international}, and countries are converging within specific 'clubs' based on structural similarities, suggesting that while IRC is increasing, it is not uniformly distributed \citep{barrios2019there}. Additionally, an earlier study by \cite{wagner2015continuing} suggested that while the global network of collaboration has become denser, it has not become more clustered, which indicates an increase in connections across the network without the formation of exclusive groups. More recent findings show that the impact of geographical distance has decreased over time (although not homogeneously), whereas the impact of political and socioeconomic distance has increased over time \citep{vieira2022distance}. Although this research provides insight on how and why certain countries might be collaborating, more systematic analyses of the effects of national factors tend to be absent.

Only a handful of studies have looked at what country-specific factors drive IRC proliferation. \cite{tsukada2015determinants} point out that the size of research teams and a nation's resource capabilities are pivotal IRC ties, while language barriers discourage IRC. Additionally, publication activities, disciplinary specialization, and shared language significantly enhance collaborative ties, which is particularly evident in China’s extensive international research networks \citep{zhang2017china}. When it comes to socio-political factors that impact IRC, mutual membership in international organizations, shared cultural or linguistic traits, and similar size of economies have positive effects on scientific collaboration \citep{hou2021impact}. Prior research also highlights differences in publication rates between larger countries, which tend to produce more domestic publications, and smaller or more remote countries, which exhibit higher rates of international publications \citep{glanzel2005domesticity}. This suggests that larger nations may rely more on their substantial domestic resources, whereas smaller countries often seek international collaboration to enhance their research impact. Gender dynamics within IRC reveal that while female scientists are more engaged in national and institutional collaborations, international collaborations are predominantly led by male scientists \citep{kwiek2021gender}. The increase in IRC is uneven across disciplines with greater acceleration in applied disciplines such as medical, social science, and geo-science, and lower levels of growth in basic disciplines such as physics and mathematics \citep{coccia2016allometric}. 

While the existing literature has significantly enhanced our understanding of the drivers and the dynamics of International Research Collaboration (IRC), it has predominantly focused on factors such as country size, resources, technological advancements, disciplinary specialization, institutional shifts \citep{tsukada2015determinants, zhang2017china, glanzel2005domesticity}, and socio-political factors, such as geopolitical affiliations and shared cultural or linguistic traits \citep{hou2021impact}. However, such studies have often overlooked the how broader political environment impacts IRC dynamics. Very few studies have explicitly focused on the impacts of national political governance on IRC. For example, \cite{Whetsell2023} shows that democratic governance is associated with the evolution of IRC ties in the 21st century. This study suggested that democratic governance may operate on IRC via the route of academic freedom but did not explicitly test for it. Building on this research, academic freedom as a corollary of democratic governance may provide a more granular explanation for IRC network evolution. 

\section{Methods}

This article employs longitudinal social network analysis methods recently developed under the family of inferential network models known as stochastic actor-oriented models (SAOM). These models are implemented in the \texttt{RSiena} package in \texttt{R}, which stands for simulated investigation for empirical network analysis \citep{Snijders2023a, Snijders2023b}. SAOMs are appropriate for analysis of network data represented as repeated network measurements over time, i.e., network panel data. These models treat individual nodes as actors that make decisions to form or dissolve ties at each time point. The method uses Monte Carlo simulation to generate significance tests on parameters of interest by implementing an objective function to model effects of behavior on tie changes and a rate function to model network evolution \citep{SnijdersPickup2016}. Inferential network models such as SAOM provide advantages over traditional methods of linear regression that do not account for dyadic dependencies in relational data \citep{cranmer2020inferential}. Even methodological innovations in dyadic econometrics, e.g., the relatively popular gravity model used for the analysis of international trade, do not account for relational biases inherent in network data.  

Data are gathered from several sources. First, network data is collected from Web of Science (WoS) Raw Data (XML) from 1993 to 2022. Research continues to support high quality assessment of WoS database \citep{Nguyen2022}. The network data are generated by parsing the XML files using \texttt{Python} to identify co-authorship instances across articles where two or more authors represent distinct country affiliations. For example, an article produced by a research team with authors from the United States (USA), the Netherlands (NLD), and China (CHN) generates three nodes, one for each affiliated country, and three ties, USA-CHN, CHN-NLD, USA-NLD. As the data are parsed, instances of country-country co-authorship stack, such that all instances of co-authorship between two countries are summed (edge weight) for each year. For example, the largest edge weight between any two countries is USA-CHN in 2022 which shows 85,482 co-authorship instances. 

To test hypothesized effects, specific network matrices were generated based on broad subject categories found in the WoS XML files: science and technology (S\&T), social sciences (SocSci), and arts and humanities (A\&H). The final resulting network data include three sets of 30 symmetrical matrices, where the rows and columns are countries and the cell values are the sum of collaborative activity between each country pair for each year. Further, to test temporal heterogeneity of affects, the networks are also analyzed in some models in three decade-long segments.

Significant country name disambiguation problems were encountered throughout the XML files. A disambiguation dictionary was compiled assigning standardized three letter ISO country codes. Ambiguous country names include, for instance, misspellings, alternative spellings, and instance where cities were listed instead of countries. GPT-4 was used to assist in this dictionary generation process. The disambiguation process started with the most current network (2022) for all articles without subject category sub-setting then moved backward to find overlooked instances until no ambiguous country names remained. After disambiguation, the largest network (2022) resulted in a matrix of 230 distinct countries including all international co-authorship instances in all article records. 

SAOMs have limited options for the analysis of continuously weighted ties. As is common in the network analysis literature, the ties are first binarized. The \texttt{R} \texttt{backbone} package developed by \citep{Neal2014} was used to transform weighted ties into binary ties. The package implements a disparity filter that identifies whether ties are statistically significant given the distribution of tie weights in the network. This procedure also trims the network such that low weight ties are re-coded as zeros and higher weight ties are preserved as ones. The procedure trimmed ties in the networks by about 85\% per year, which also resulted in disconnection of nodes from approximately 44\% in 1993 to less than 5\% in 2022. The conceptual interpretation given the implementation of the procedure is that only high tie strength 'significant' connections remain in the networks. Thus, effects on network ties can be understood in terms of tie strength despite the lack of tie weight information present after binarizing. 

The primary independent variable of interest is the Academic Freedom Index \citep{Spannagel2023}, which was gathered from the \texttt{R} package \texttt{vdemdata} that contains data from the Varieties of Democracy Project \citep{Coppedge2011}. The Varieties of Democracy Project \citep{VDEM2024} defines academic freedom in the following manner.

    \begin{quote}“Academic freedom is understood as the right of academics, without constriction by prescribed doctrine, to freedom of teaching and discussion, freedom in carrying out research and disseminating and publishing the results thereof, freedom to express freely their opinion about the institution or system in which they work, freedom from institutional censorship and freedom to participate in professional or representative academic bodies.” \citep{VDEM2024}\end{quote}  
    
The Academic Freedom Index is a continuous variable that ranges between zero and one. The index is an aggregate of empirical sub-measures. It is created from Bayesian factor analysis, conducted on five sub-indexes: freedom of teaching/research, academic exchange/dissemination, institutional autonomy, campus integrity, and freedom of expression \citep{VDEM2024,Pemstein2023}. The Academic Freedom Index is available for 178 countries from 1900 to 2022 with increasing data coverage up to the latest year. 

While existing research has tested the effects of democracy on IRC \citep{Whetsell2023}, academic freedom represents a more granular construct that is closer to IRC in the theoretical chain of causation. The preliminary test of the effect of liberal democracy on academic freedom shows a bivariate correlation of 0.84, while a two-way fixed effects model on unbalanced panel data using \texttt{plm} showed a very similar estimate of 0.84 with a standard error of 0.012 and an adjusted R-square of 0.5 (n=4965) for the 166 countries during the 30-year time frame.

For the SAOMs, the Academic Freedom Index is included as a direct effect (egoPlusAtlX) on network evolution and as a homophily effect (simX). Since academic freedom is the primary variable of interest, network nodes without data on this variable were dropped. Countries that did not continuously exist across the time frame were dropped, resulting in a uniform node set of 166 countries across 30 years from 1993 to 2022. Several control variables were included to account for potential confounders. To account for country-level resources, the World Bank's GDP/PPP measure is used. This variable has adequate data coverage and is comparable across countries. International research collaboration is known to be affected by country size \citep{Melin1999}, which can be measured in a number of ways. As such, population size is included in the model. The level of urbanization was also included, as it has been shown to affect country level innovation \citep{wang2021impacts}. GDP, population size, and urbanization were gathered from the \texttt{R} package \texttt{WDI}. Finally, it is well known that international research collaboration tends to decrease with geographic distance \citep{Katz1994}. Thus, the model includes a geographical distance matrix as a constant dyadic covariate (coDyadvar), accounting for the network evolution effects of distance. The distance data was generated from the \texttt{cshapes} package in \texttt{R}.

Measures of scientific capacity such as the World Bank's RDGDP or the OECD's GERD are not included because they have limited data coverage which drastically reduces the sample size to mostly developed countries. There is partial missing GDP/PPP data for some countries, and completely missing GDP/PPP data for CUB, PRK, SYR, and TWN. In total 4.9\% of observations were missing. \texttt{RSiena} imputes the global average for missing data during the simulation phase of estimation but imputed data are not used to generate target statistics in the SAOMs \citep{Snijders2023a}.  

Apart from geographic distance, all exogenous controls are included as direct effects (egoPlusAltX) and homophily effects (simX). Geographical distance, GDP/PPP, and population size were log-transformed (+1) for use in the models. Finally, three endogenous network controls are included in the model: a term that models the density (degree) of the network, a term that models the distribution of triads in the network is included (gwesp), and  a term that models the degree distribution in undirected networks is included (degPlus).

\section{Results}

The results section is organized as follows. First, descriptive statistics on the networks over time are presented. Next, a visualization of the S\&T network is shown. Next, the results of the stochastic actor oriented models (SAOM) showing heterogeneous effects across research domains for the full time frame are presented. Lastly, models are shown for each research domain that show heterogeneous effects in consecutive temporal segments. 

Table 1 shows the number of nodes, number of edges, and the density for the science and technology network (S\&T), the social sciences network (SocSci), and the arts and humanities network (A\&H). For all three networks, the statistics are calculated after network trimming using the disparity filter from the \texttt{R} package \texttt{backbone} \citep{Neal2014}. Hence, the networks are interpreted as including only 'significant' ties. The S\&T network is much larger by number of edges and density than SocSci or A\&H. In particular the A\&H network is a virtually non-existent in the early years. The number of nodes is uniform for all three node sets in order to preserve comparability between them and because \texttt{RSiena} requires a uniform node set. 

\begin{table}[H]
\centering
\caption{Network Descriptive Statistics for S\&T, SocSci, and A\&H}
\small
\begin{adjustbox}{center}
\begin{tabular}{lccccccc}
\hline
Year & Nodes & \multicolumn{3}{c}{Edges} & \multicolumn{3}{c}{Density} \\
     &       & S\&T & SocSci & A\&H & S\&T & SocSci & A\&H \\
\hline
1993 & 166	&289	&35	    &2	    &0.021	&0.003	&0.000\\
1994 & 166	&324	&57	    &1	    &0.024	&0.004	&0.000\\
1995 & 166	&367	&53	    &2	    &0.027	&0.004	&0.000\\
1996 & 166	&365	&52	    &2	    &0.027	&0.004	&0.000\\
1997 & 166	&426	&58	    &4	    &0.031	&0.004	&0.000\\
1998 & 166	&456	&87	    &10	    &0.033	&0.006	&0.001\\
1999 & 166	&459	&115	&9	    &0.034	&0.008	&0.001\\
2000 & 166	&479	&104	&12	    &0.035	&0.008	&0.001\\
2001 & 166	&479	&86	    &6	    &0.035	&0.006	&0.000\\
2002 & 166	&510	&93	    &14	    &0.037	&0.007	&0.001\\
2003 & 166	&545	&141	&14	    &0.040	&0.010	&0.001\\
2004 & 166	&592	&152	&11	    &0.043	&0.011	&0.001\\
2005 & 166	&609	&125	&16	    &0.044	&0.009	&0.001\\
2006 & 166	&665	&226	&30	    &0.049	&0.017	&0.002\\
2007 & 166	&706	&180	&40	    &0.052	&0.013	&0.003\\
2008 & 166	&790	&253	&44	    &0.058	&0.018	&0.003\\
2009 & 166	&826	&285	&57	    &0.060	&0.021	&0.004\\
2010 & 166	&873	&263	&64	    &0.064	&0.019	&0.005\\
2011 & 166	&840	&284	&67	    &0.061	&0.021	&0.005\\
2012 & 166	&933	&299	&73	    &0.068	&0.022	&0.005\\
2013 & 166	&948	&338	&98	    &0.069	&0.025	&0.007\\
2014 & 166	&1104	&375	&99	    &0.081	&0.027	&0.007\\
2015 & 166	&1200	&406	&107	&0.088	&0.030	&0.008\\
2016 & 166	&1280	&478	&146	&0.093	&0.035	&0.011\\
2017 & 166	&1330	&486	&148	&0.097	&0.035	&0.011\\
2018 & 166	&1355	&531	&148	&0.099	&0.039	&0.011\\
2019 & 166	&1571	&564	&156	&0.115	&0.041	&0.011\\
2020 & 166	&1630	&574	&154	&0.119	&0.042	&0.011\\
2021 & 166	&1643	&590	&157	&0.120	&0.043	&0.011\\
2022 & 166	&1632	&566	&142	&0.119	&0.041	&0.010\\
\hline
\end{tabular}
\end{adjustbox}
\end{table}

Figure 1 illustrates the S\&T network evolution across the 30-year time frame. Each network cross-section is labelled by year. The nodes are sized by node degree. The color of the nodes is taken from the Academic Freedom Index for the node in the given year. From visual inspection it is evident that the network dramatically increases in size and density. Also, it appears that high academic freedom nodes (blue) remain central throughout, but that low academic freedom nodes (red) also appear to become more central in the network over time.

\begin{figure}[H]
\caption{Visualization of S\&T Network Evolution, 1993-2022}
\centering
\includegraphics[scale= 0.4]{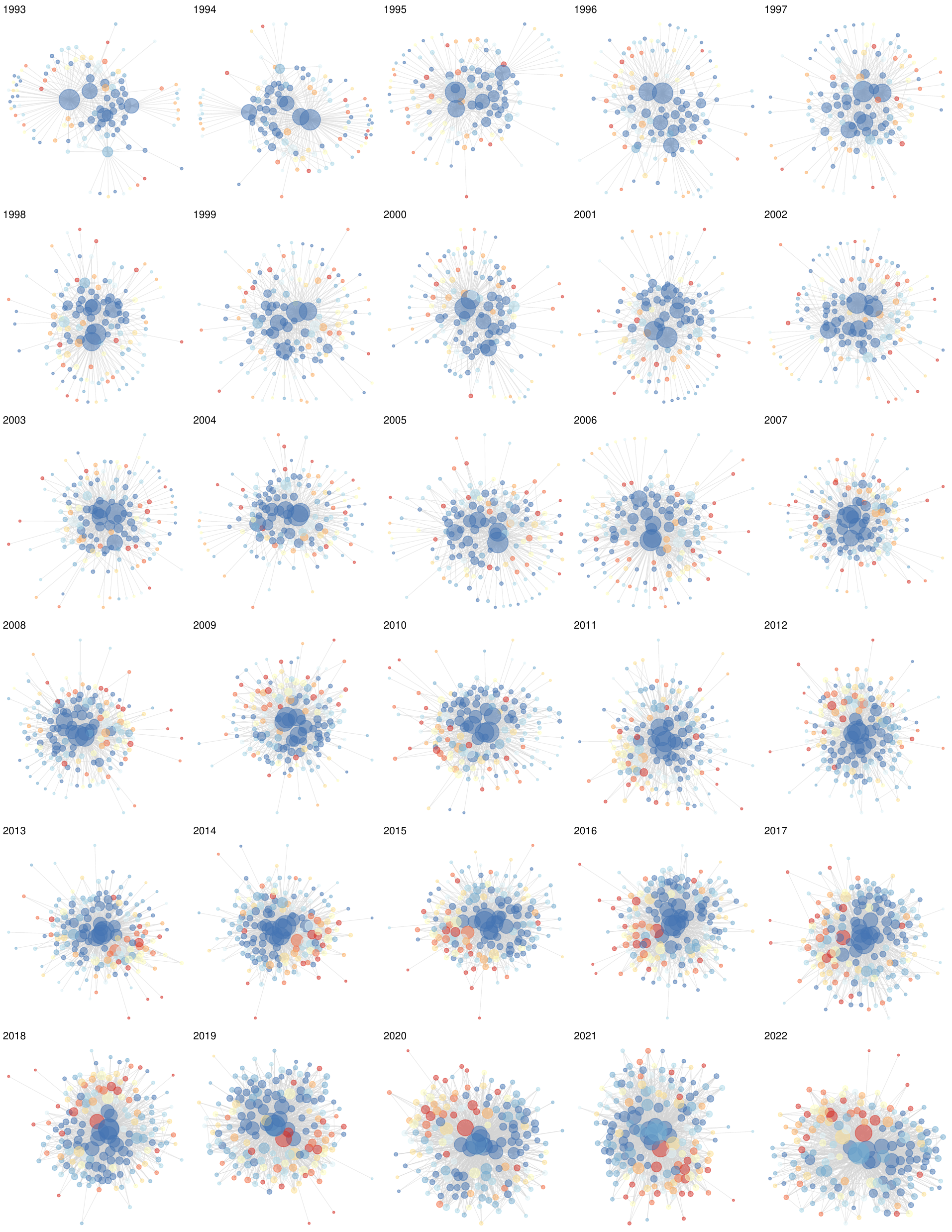} 
\end{figure}

Next, Table 2 shows the results of the stochastic actor oriented models (SAOM) executed with \texttt{RSiena}. The model estimates indicate the sign, strength, and significance of the parameter estimates on the formation and maintenance of strong international research collaboration ties, i.e. network evolution. Standard errors are included in parentheses under the estimates. There are three models, one for each domain of research: S\&T, SocSci, and A\&H. The table is divided by primary independent variables at the top, exogenous control variables in the middle, and endogenous controls at the bottom. 

\begin{table}[H]
\centering
\caption{SAOMs, IRC Evolution, 1993-2022}
\begin{tabular}{lccc}
\hline
Parameter & S\&T & SocSci & A\&H \\
\hline
free (egoPlusAtlX)       &  0.5152       & 0.9968        & 1.9003       \\
                            & (0.0451)      & (0.0779)      & (0.2232)     \\
free (simX)        &  0.4441       & 0.2670        & -0.0343      \\
                            & (0.0651)      & (0.1049)      & (0.2595)     \\
\hline                            
gdp (egoPlusAtlX)        &  0.1492       & 0.4848        & 0.8728       \\
                            & (0.0231)      & (0.0402)      & (0.0762)     \\
gdp (simX)         & -2.3148       & -2.1915       & 2.1969       \\
                            & (0.1903)      & (0.3236)      & (0.6678)     \\
pop (egoPlusAtlX)        &  0.1318       & -0.0283       & -0.2263      \\
                            & (0.0193)      & (0.0334)      & (0.0666)     \\
pop (simX)         &  1.0637       & 1.5914        & -0.7173      \\
                            & (0.1372)      & (0.2269)      & (0.4607)     \\
urb (egoPlusAtlX)        &  0.0054       & 0.0064        & 0.0072       \\
                            & (0.0009)      & (0.0014)      & (0.0028)     \\
urb (simX)         & -0.2309       & -0.3251       & 0.8163       \\
                            & (0.0814)      & (0.1355)      & (0.3166)     \\
geo. dist. (coDyadCovar)             & -1.1125       & -0.7854       & -0.5624      \\
                            & (0.0211)      & (0.0279)      & (0.0469)     \\
\hline                            
Density (degree)       &  -5.0028      & -6.5302       & -9.1706      \\
                            & (0.0869)      & (0.0848)      & (0.1748)     \\
Transitivity (gwesp)              &  0.8727       & 1.0502        & 0.7503       \\
                            & (0.0506)      & (0.0506)      & (0.0827)     \\
Pref. Attach (degPlus)         &  0.0219       & 0.0273        & 0.0507       \\
                            & (0.0008)      & (0.0012)      & (0.0041)     \\                            
\hline
Convergence Ratio           &  0.0690       & 0.0867        & 0.0993       \\
Iteration Steps             &  3974         & 3832          & 3837         \\
\hline
\end{tabular}
\vspace{2.5pt} 
\parbox{0.8\textwidth}{\footnotesize Note: Standard errors in parentheses. Convergence ratios > 0.25 = poor fit, 0.1 - 0.25 = acceptable fit, < 0.1 = excellent fit}
\end{table}

The direct effect (egoPlusAtlX) of the Academic Freedom Index (free) is positive and significant across all three models. As expected, the size of the estimates increase from S\&T to SocSci and is strongest in A\&H. This is consistent with the hypothesis suggesting academic freedom is more salient in potentially politically sensitive domains. Finally, the homophily term (simX) follows the opposite pattern, decreasing from S\&T to SocSci to A\&H. This is perhaps because the direct effects of academic freedom are stronger than homophily effects in more politically sensitive domains. However, it also may be the case that researchers are more likely to collaborate with countries with similar academic freedom levels in S\&T and SocSci than in A\&H where the term is non-significant. 

The exogenous controls show interesting results. GDP appears to have a positive and significant estimate in all three domains, while GDP homophily appears negative and significant in S\&T and SocSci, and the reverse in A\&H. Population size appears to show almost the opposite pattern for direct and homophily effects. Urbanization shows a positive significant effect, except in A\&H, with the opposite homophily pattern. Geographic distance shows negative and significant estimates in all three models, indicating strong ties are less likely when distance increases.

The endogenous controls capturing general network processes show density (degree) has a negative sign indicating relatively low density, where S\&T is the most dense, then SocSci, and A\&H is the least dense. Transitivity (gwesp) models the distribution of transitive triads in the network, and the estimates are positive and significant in all three models indicating that transitivity is an operative process in the networks. Pref.Attach (degPlus) models the degree distribution suggesting that a preferential attachment process is present, where nodes have a tendency to connect preferentially to already strongly connected nodes in the networks. The convergence ratios show excellent fits for each model, where below 0.1 is considered excellent, between 0.1 and 0.25 is acceptable, and above 0.25 is poor. Diagnostic plots are shown in the supplementary information section of the paper.

Next, given the visual inspection of Figure 1, the effect of academic freedom on IRC appears to be temporally heterogeneous, i.e., that the effect is weakening over time as autocratic countries are penetrating the network. Further, the direct effect of academic freedom may be overpowered by the homophily effect between countries. 

Table 3 shows SAOMs for three distinct time periods for the S\&T network. As the models show, the effect of academic freedom on IRC indeed appears to be weakening, with the strongest effect from 1993 to 2002, a weaker though still significant effect from 2003 to 2012, and the weakest and now insignificant effect from 2013 to 2022. Concurrently, the homophily effect appears to show the opposite pattern, which suggests a greater sorting of strong IRC ties between countries with similar levels of academic freedom. 

\begin{table}[H]
\centering
\caption{SAOMs, Science and Technology, Temporal Heterogeneity}
\begin{tabular}{lccc}
\hline
Parameter & 1993:2002 & 2003:2012 & 2013:2022 \\
\hline
free (egoPlusAtlX)        &  0.6816   &  0.2519   &  0.0510   \\
                            & (0.1643)  & (0.0905)  & (0.0597) \\
free (simX)         &  0.2679   &  0.4481   &  0.8420   \\
                            & (0.1947)  & (0.1191)  & (0.0873) \\
\hline                            
gdp (egoPlusAtlX)         &  0.3506   &  0.3095   &  0.0608   \\
                            & (0.0804)  & (0.0472)  & (0.0313) \\
gdp (simX)          & -2.4249  & -1.4039  & -1.6419  \\
                            & (0.5784)  & (0.3097)  & (0.2021) \\
pop (egoPlusAtlX)         &  0.0688   & -0.0400   &  0.0926   \\
                            & (0.0646)  & (0.0400)  & (0.0277) \\
pop (simX)          &  1.9804   &  1.3637   &  0.2677   \\
                            & (0.3730)  & (0.2340)  & (0.1659) \\
urb (egoPlusAtlX)         &  0.0137   &  0.0009   &  0.0006   \\
                            & (0.0027)  & (0.0018)  & (0.0013) \\
urb (simX)          & -1.3324  & -0.7618  &  0.1314   \\
                            & (0.2261)  & (0.1471)  & (0.1087) \\
geo. dist. (coDyadCovar)              & -1.0706  & -1.1011  & -1.1419  \\
                            & (0.0519)  & (0.0395)  & (0.0298) \\  
\hline                            
Density (degree)        & -6.0517  & -4.7905  & -6.3973  \\
                            & (0.1674)  & (0.1530)  & (0.2536) \\
Transitivity (gwesp)               &  0.8444   &  0.6892   &  1.7138   \\
                            & (0.1054)  & (0.0916)  & (0.1281) \\
Pref. Attach (degPlus)          &  0.0297   &  0.0249   &  0.0238   \\
                            & (0.0028)  & (0.0015)  & (0.0011) \\                            
\hline
Convergence Ratio           &  0.0776 & 0.0683 & 0.0802 \\
Iteration Steps             & 3691 & 3803 & 4046 \\
\hline
\end{tabular}
\vspace{2.5pt}
\parbox{0.8\textwidth}{\footnotesize Note: Standard errors in parentheses. Convergence ratios > 0.25 = poor fit, 0.1 - 0.25 = acceptable fit, < 0.1 = excellent fit}
\end{table}

Table 4 shows the models for SocSci in distinct three consecutive time periods. The models show a similar pattern observed in Table 3 with a declining direct affect of academic freedom on IRC and an increasing homophily effect that gains significance in the last period. 

\begin{table}[H]
\centering
\caption{SAOMs, Social Sciences, Temporal Heterogeneity}
\begin{tabular}{lccc}
\hline
Parameter & 1993:2002 & 2003:2012 & 2013:2022 \\
\hline
free (egoPlusAtlX)        &  2.0112  &  0.7516  &  0.0459  \\
                            & (0.6611)  & (0.2033)  & (0.0865) \\
free (simX)         &  0.9142  &  0.1075  &  0.4872  \\
                            & (0.7325)  & (0.2346)  & (0.1170) \\
\hline                            
gdp (egoPlusAtlX)         &  1.1962  &  0.6965  &  0.2059  \\
                            & (0.1706)  & (0.0822)  & (0.0455) \\
gdp (simX)          &  2.5148  & -1.5130  & -0.9746  \\
                            & (1.2813)  & (0.5668)  & (0.3070) \\
pop (egoPlusAtlX)         & -0.4652  & -0.2673  &  0.0211  \\
                            & (0.1425)  & (0.0716)  & (0.0387) \\
pop (simX)          &  0.3905  &  1.7433  &  0.8534  \\
                            & (0.8930)  & (0.3974)  & (0.2243) \\
urb (egoPlusAtlX)         &  0.0078  &  0.0022  &  0.0026  \\
                            & (0.0062)  & (0.0032)  & (0.0018) \\
urb (simX)          & -2.0247  & -0.9751  & -0.1629  \\
                            & (0.6781)  & (0.2812)  & (0.1513) \\
geo. dist. (coDyadCovar)              & -0.5702  & -0.6481  & -0.8374  \\
                            & (0.1045)  & (0.0551)  & (0.0355) \\
\hline                            
Density (degree)        & -10.4599 & -7.1339  & -5.5953  \\
                            & (0.4381)  & (0.1759)  & (0.1373) \\
Transitivity (gwesp)               &  0.5520  &  1.2451  &  1.2034  \\
                            & (0.2290)  & (0.1064)  & (0.0810) \\
Pref. Attach (degPlus)          &  0.0758  &  0.0401  &  0.0318  \\
                            & (0.0093)  & (0.0029)  & (0.0016) \\                            
\hline
Convergence Ratio           &  0.1085  &  0.1231  &  0.0820  \\
Iteration Steps             &  3693    &  3460    &  3460    \\
\hline
\end{tabular}
\vspace{2.5pt}
\parbox{0.8\textwidth}{\footnotesize Note: Standard errors in parentheses. Convergence ratios > 0.25 = poor fit, 0.1 - 0.25 = acceptable fit, < 0.1 = excellent fit}
\end{table}

Lastly, Table 5 shows a decreasing direct effect (significant only in the final time period) but non-significant homophily effects in all three time periods. Notably, the convergence statistics in the first two time period models are very poor, reflecting the extreme sparsity of these networks in the early periods. Thus, only the final model here contains reliable estimates. 

\begin{table}[H]
\centering
\caption{SAOMs, Arts and Humanities, Temporal Heterogeneity}
\begin{tabular}{lccc}
\hline
Parameter & 1993:2002 & 2003:2012 & 2013:2022 \\
\hline
free (egoPlusAtlX)        &  2.9625  &  2.5072  &  1.0715 \\
                            & (3.6201) & (0.6844) & (0.2717) \\
free (simX)         &  3.8063  &  0.9006  & -0.2267 \\
                            & (4.5468) & (0.7849) & (0.3055) \\
\hline                            
gdp (egoPlusAtlX)         &  2.4641  &  1.2688  &  0.4816 \\
                            & (0.5781) & (0.1479) & (0.0844) \\
gdp (simX)          &  4.7304  &  5.6474  &  2.1303 \\
                            & (5.1744) & (1.1767) & (0.6385) \\
pop (egoPlusAtlX)         & -0.7199  & -0.5394  & -0.0734 \\
                            & (0.5712) & (0.1379) & (0.0709) \\
pop (simX)          & -7.1700  & -3.2754  & -0.3677 \\
                            & (4.8046) & (0.9090) & (0.4425) \\
urb (egoPlusAtlX)         &  0.0185  &  0.0032  &  0.0049 \\
                            & (0.0251) & (0.0062) & (0.0033) \\
urb (simX)          &  1.7687  & -0.3968  &  0.1673 \\
                            & (3.2704) & (0.7029) & (0.3182) \\
geo. dist. (coDyadCovar)              & -0.5424  & -0.4156  & -0.6122 \\
                            & (0.3345) & (0.0945) & (0.0597) \\
\hline                            
Density (degree)        & -22.5824 & -11.1655 & -7.1684 \\
                            & (2.6932) & (0.4348) & (0.1495) \\
Transitivity (gwesp)               & -0.1696  &  0.3108  &  0.9418 \\
                            & (0.9692) & (0.1957) & (0.1227) \\
Pref. Attach (degPlus)          &  0.0593  &  0.0951  &  0.0607 \\
                            & (0.1177) & (0.0136) & (0.0057) \\                            
\hline
Convergence Ratio           &  0.8934  &  1.0994  &  0.0968 \\
Iteration Steps             &  3332    &  3242    &  3242   \\
\hline
\end{tabular}
\vspace{2.5pt}
\parbox{0.8\textwidth}{\footnotesize Note: Standard errors in parentheses. Convergence ratios > 0.25 = poor fit, 0.1 - 0.25 = acceptable fit, < 0.1 = excellent fit}
\end{table}

\section{Discussion}

The findings of this study highlight the important role that academic freedom plays in creating the conditions for the flourishing of international research. The positive and significant estimates observed across different domains reinforce the hypothesis that academic freedom provides the conditions for free association across political boundaries. Heterogeneous effects across disciplines, more pronounced in the social sciences and especially in the arts and humanities, suggest that academic freedom is critical to maintaining global research collaboration networks, particularly in politically salient disciplines. 

However, the results also highlight a worrying trend. Academic freedom appears to be becoming more and more disassociated with IRC over in recent years, where the direct effect of academic freedom on IRC is supplanted by an assortative pattern where countries of similar academic freedom appear to be developing strong ties. In other words, democracies collaborate with democracies and autocracies collaborate with autocracies. This finding conforms to the most restrictive tie strength model presented in \cite{Whetsell2023}.  

Although autocratic regimes tend to impose constraints on intellectual freedom, we have shown that China is now a significant player in the IRC network, promoting global collaboration even as it maintains strict control over academic institutions. Similarly, former post-Soviet republics that have varying degrees of autocracy have engaged actively in IRC. Although these examples illustrate that autocratic regimes perhaps stimulate IRC as a strategy to overcome domestic limitations, the findings here suggest that countries with higher levels of academic freedom continue to be more strongly associated with strong IRC evolution over a longer time span. 

Further, while recent research demonstrated positive effects of democratic governance on international research collaboration for a more limited time frame \citep{Whetsell2023}, the current study provides a more granular approach in a couple of key ways. First, this study leverages the recently developed Academic Freedom Index, which is closer than democracy in the causal chain between political governance and cross-border collaborative activity. Second, the current study examines a much longer stretch of time from the early 1990s to the 2020s. Third, the current study provides estimates on distinct research domains. While it of course remains difficult to ascribe causality without intervention or randomization, the current study leverages the latest advancements in social network analysis \citep{Snijders2023a} and a large corpus of high quality data in both the primary independent \citep{Spannagel2023} and dependent variable \citep{Nguyen2022}. 

Lastly, this study suggests a potential explanation for recent observations concerning a slow down in scientific innovation \citep{park2023papers}. Previous research has demonstrated that IRC tends to correspond to conventional rather than novel research \citep{wagner2019international}. Increasing penetration of the global network of science by autocratic countries may be resulting in less disruptive research overall. However, supporting this conjecture requires further testing in a way that empirically connects academic freedom, IRC, and disruptiveness. 

\section{Conclusion}

The primary purpose of this research has been to address high level questions about the effects of political processes on the function of science in the international system. In accordance with longstanding conceptual analysis and a small amount of recent empirical work, we provide global scale empirical evidence over a thirty year time frame to show that academic freedom indeed appears to support international research collaboration. The effect increases in research domains with a greater preponderance of politically salient disciplines. The strongest effects were observed in the arts and humanities, followed by social sciences, and then science and technology. This could mean that disciplines more susceptible to political and ideological influences benefit most from academic freedom. In fields like arts and humanities, where research often intersects with cultural and political issues, academic freedom ensures that scholars can engage in critical and innovative work without fear of censorship or retribution.

The current landscape of academic freedom is marked by both challenges and opportunities. Democratic backsliding and rising authoritarianism in various parts of the world pose significant threats to academic freedom, as institutions of higher education become targets for autocratic control. Our findings align with prior literature that emphasizes the protective role of academic freedom in democratic societies, where it not only fosters innovation and knowledge production but also upholds broader democratic values and societal well-being. In sum, greater international research collaboration appears to be empirically dependent on an atmosphere of academic freedom in which researchers are free to associate across political borders in order to pursue the aims of scholastic curiosity. However, this research also documents a concerning trend in recent times. As autocratic countries play an increasingly important role in global science, academic freedom appears to be increasingly disassociated with processes of international collaboration. Rather, countries appear to be sorting into collaborative patterns that preferentially select for similar levels of freedom and repression.

\bibliographystyle{apacite}

\bibliography{bib}

\appendix

\section{Appendix}

\subsection{Models Excluding USA and CHN}

Since USA and CHN are so productive and prolific in the network, there are some concerns that inclusion may skew the results. To alleviate this concern, we first excluded both from the WoS XML parsing such that all records including USA or CHN would be ignored in the matrix generation process. Compared to Model 1 in Table 2, the results are very similar. Thus, it appears the results are not sensitive to the exclusion of USA and CHN. 

\begin{table}[H]
\centering
\caption{SAOMs, Science and Technology, 1993-2022}
\begin{tabular}{lcc}
\hline
free (egoPlusAtlX)       & 0.4650    \\ 
& (0.0426)       \\
free (simX)        & 0.4645     \\
& (0.0624)       \\
gdp (egoPlusAtlX)        & 0.0836     \\
& (0.0198)       \\
gdp (simX)         & -1.7879    \\
& (0.1527)       \\
pop (egoPlusAtlX)        & 0.1633     \\
& (0.0175)       \\
pop (simX)         & 0.9476     \\
& (0.1247)       \\
urb (egoPlusAtlX)        & 0.0060     \\
& (0.0008)       \\
urb (simX)         & -0.1949    \\
& (0.0744)       \\
geo. dist. (coDyadCovar)             & -1.0749    \\
& (0.0194)       \\
Density (degree)           & -4.6491    \\
& (0.0676)       \\
Transitivity (gwesp)                  & 0.8232     \\
& (0.0416)       \\
Pref. Attach (degPlus)             & 0.0214     \\
& (0.0007)       \\
\hline
Overall max convergence ratio   & 0.0552                      \\
Total iteration steps           & 4005                       \\
\hline
\end{tabular}
\label{table:saom-parameters}
\end{table}

\subsection{Goodness of Fit Plots}

The top row of figures shows the fit for the degree distributions for each network. Subplot (a) shows some difficulty in modeling nodes with zero edges. The second row shows the triad census for each network. Subplot (e) and (f) show some difficulty in modeling triads of type 300, where all three nodes are connected. This may be due to the high number of such triads resulting from the undirected nature of connections in multi-authored papers.

\begin{figure}[H]
    \centering
    \caption{Goodness of Fit Plots for SAOMs}
    \begin{subfigure}{.3\textwidth}
        \centering
        \includegraphics[width=\linewidth]{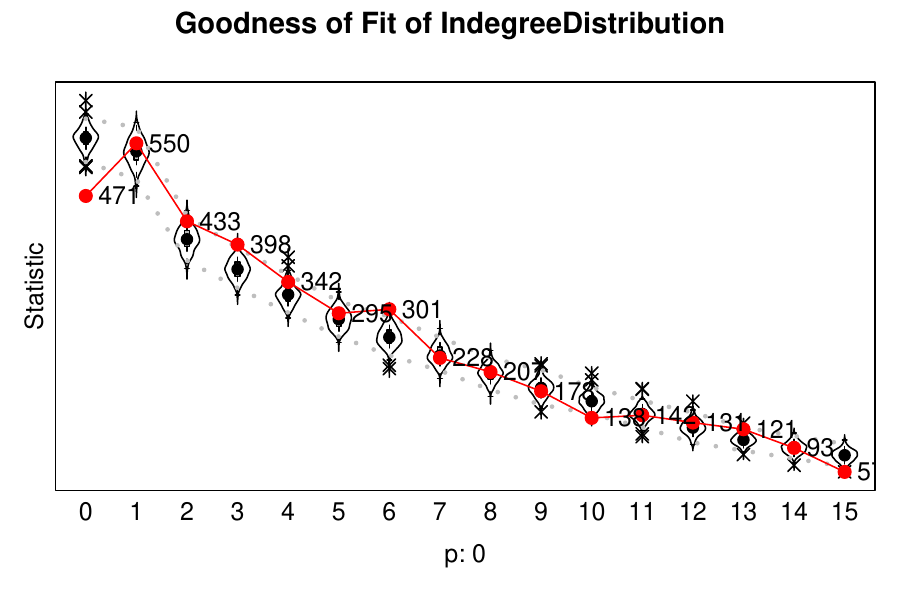}
        \caption{S\&T}
    \end{subfigure}\hfill
    \begin{subfigure}{.3\textwidth}
        \centering
        \includegraphics[width=\linewidth]{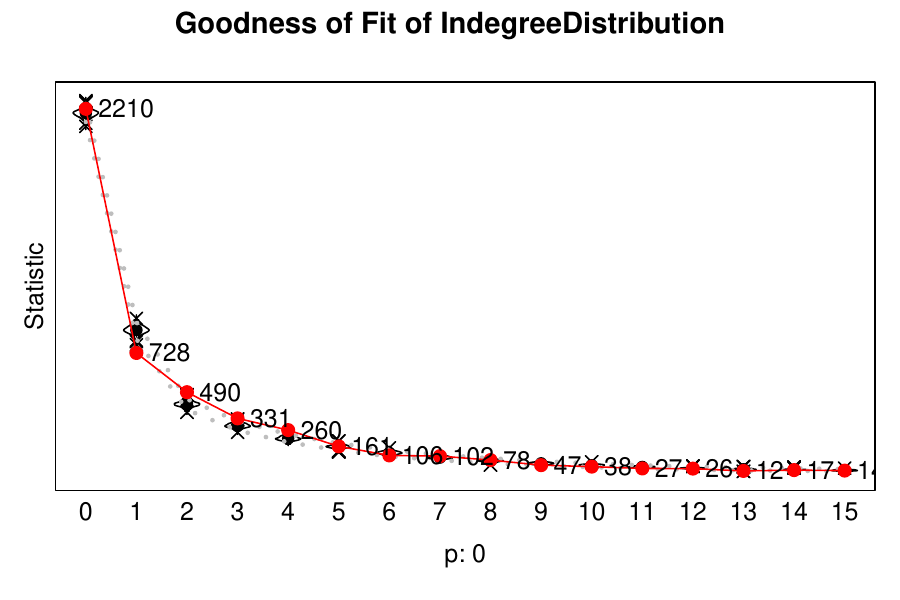}
        \caption{SocSci}
    \end{subfigure}\hfill
    \begin{subfigure}{.3\textwidth}
        \centering
        \includegraphics[width=\linewidth]{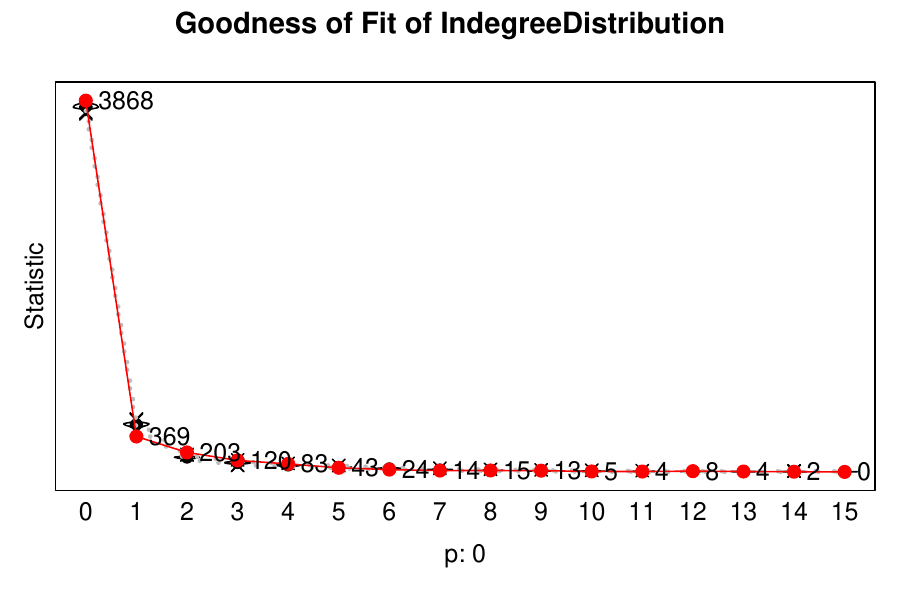}
        \caption{A\&H}
    \end{subfigure}
    \medskip
    \begin{subfigure}{.3\textwidth}
        \centering
        \includegraphics[width=\linewidth]{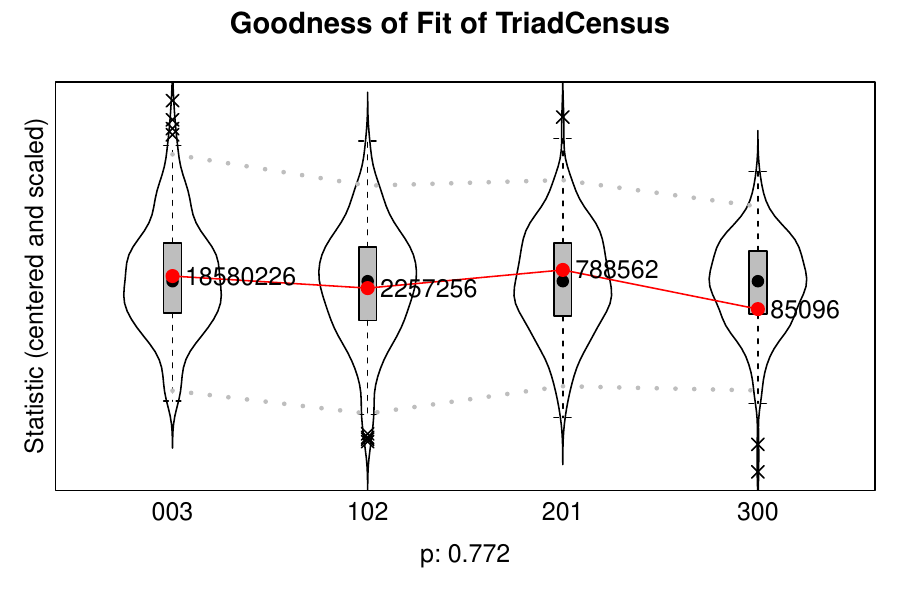}
        \caption{S\&T}
    \end{subfigure}\hfill
    \begin{subfigure}{.3\textwidth}
        \centering
        \includegraphics[width=\linewidth]{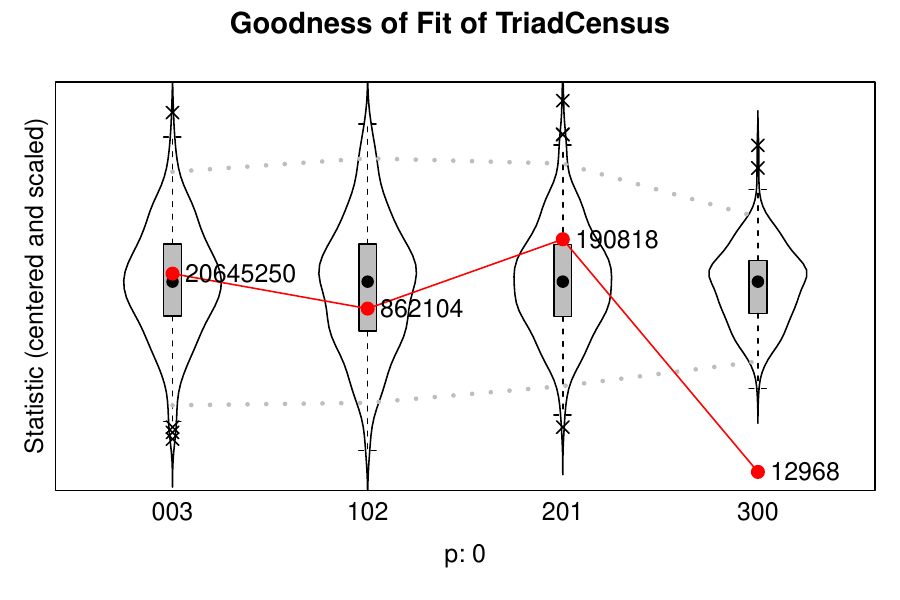}
        \caption{SocSci}
    \end{subfigure}\hfill
    \begin{subfigure}{.3\textwidth}
        \centering
        \includegraphics[width=\linewidth]{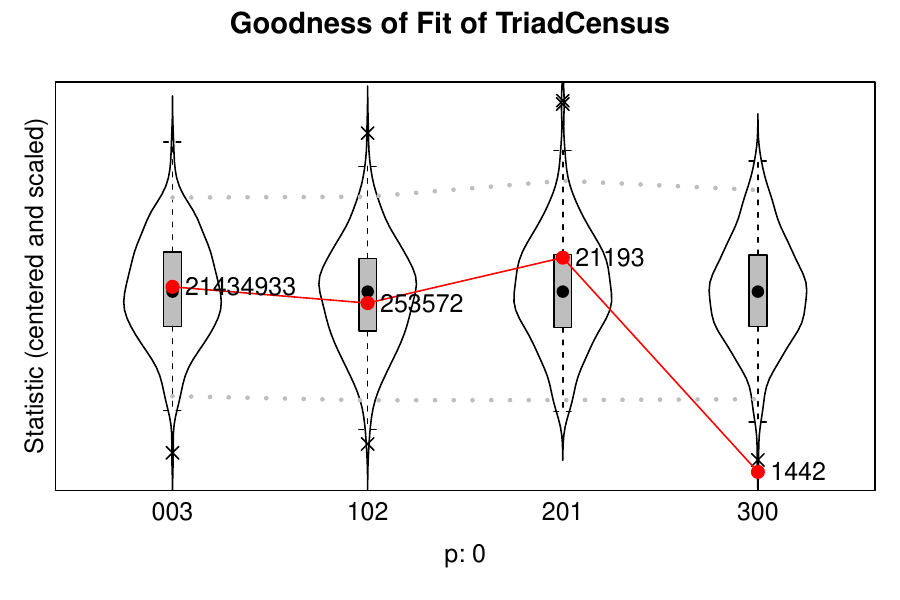}
        \caption{A\&H}
    \end{subfigure}
    \captionsetup{font=small}
    \caption*{Note: These figures illustrate the goodness of fit plots for SAOMs in various disciplines.}
\end{figure}

\end{document}